\documentclass[conference]{IEEEtran}
\IEEEoverridecommandlockouts
\usepackage{cite}
\usepackage{amsmath,amssymb,amsfonts}
\usepackage{algorithm}      
\usepackage{algorithmic}    
\usepackage{graphicx}
\usepackage{textcomp}
\usepackage{xcolor}

\def\BibTeX{{\rm B\kern-.05em{\sc i\kern-.025em b}\kern-.08em
    T\kern-.1667em\lower.7ex\hbox{E}\kern-.125emX}}
\begin{document}

\title{Hybrid Quantum–Classical Policy Gradient for Adaptive Control of Cyber-Physical Systems: A Comparative Study of VQC vs. MLP\\

}

\author{
\IEEEauthorblockN{Aueaphum Aueawatthanaphisut\textsuperscript{*} and Nyi Wunna Tun}
\IEEEauthorblockA{
\textit{School of Information, Computer, and Communication Technology} \\
\textit{Sirindhorn International Institute of Technology, Thammasat University}\\
Pathum Thani, Thailand \\
Email: aueawatth.aue@gmail.com, 6722790258@g.siit.tu.ac.th
\\[2pt]
\textit{\footnotesize *Corresponding Author: Aueaphum Aueawatthanaphisut}
}
}

\maketitle

\begin{abstract}
The comparative evaluation between classical and quantum reinforcement learning (QRL) paradigms was conducted to investigate their convergence behavior, robustness under observational noise, and computational efficiency in a benchmark control environment. The study employed a multilayer perceptron (MLP) agent as a classical baseline and a parameterized variational quantum circuit (VQC) as a quantum counterpart, both trained on the CartPole-v1 environment over 500 episodes. Empirical results demonstrated that the classical MLP achieved near-optimal policy convergence with a mean return of $498.7 \pm 3.2$, maintaining stable equilibrium throughout training. In contrast, the VQC exhibited limited learning capability, with an average return of $14.6 \pm 4.8$, primarily constrained by circuit depth and qubit connectivity. 

Noise robustness analysis further revealed that the MLP policy deteriorated gracefully under Gaussian perturbations, while the VQC displayed higher sensitivity at equivalent noise levels. Despite the lower asymptotic performance, the VQC exhibited significantly lower parameter count and marginally increased training time, highlighting its potential scalability for low-resource quantum processors. The results suggest that while classical neural policies remain dominant in current control benchmarks, quantum-enhanced architectures could offer promising efficiency advantages once hardware noise and expressivity limitations are mitigated.
\end{abstract}

\begin{IEEEkeywords}
Quantum Reinforcement Learning, Variational Quantum Circuit, CartPole-v1, Classical vs Quantum Comparison, Noise Robustness, Convergence Stability, Computational Efficiency
\end{IEEEkeywords}

\section{Introduction}
Reinforcement learning (RL) has emerged as one of the central paradigms for sequential decision making, enabling autonomous agents to learn control strategies through interaction with their environments. Classical RL algorithms such as Q-learning and policy gradient methods have achieved remarkable success in robotics, autonomous driving, and cyber–physical control systems. Nevertheless, their scalability is often hindered by the curse of dimensionality and slow convergence in complex, nonlinear environments.  

\begin{figure}
    \centering
    \includegraphics[width=1\linewidth]{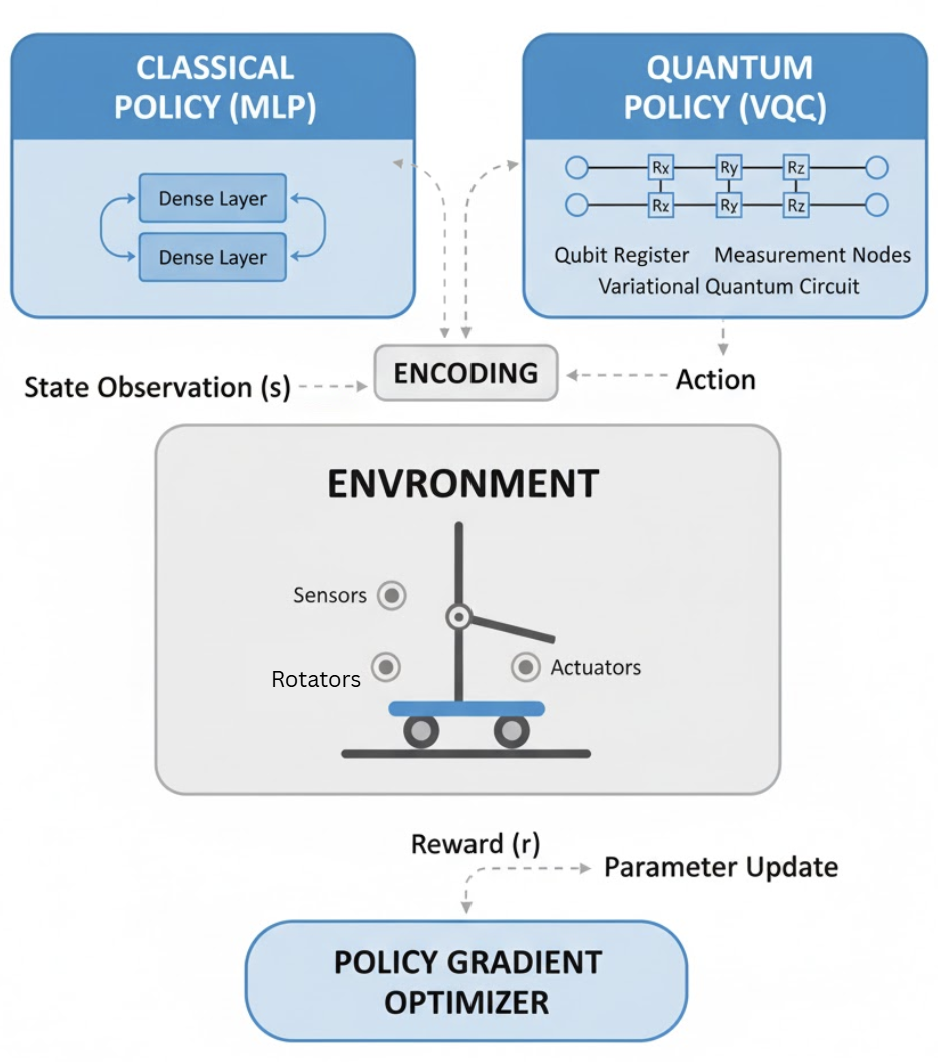}
    \caption{Framework of Hybrid Quantum-Classical Policy Gradient Reinforcement Learning to integrate a Variational Quantum Circuit (VQC) and a Classical MLP.}
    \label{fig:placeholder}
\end{figure}

In recent years, the intersection of quantum computing and machine learning has given rise to a new class of algorithms—quantum reinforcement learning (QRL)—that seek to exploit quantum mechanical principles such as superposition, entanglement, and quantum parallelism to enhance exploration efficiency and learning speed. The foundational framework of QRL was first proposed by Dong \textit{et al.} [1], where quantum states were used to represent policy superpositions and measurement collapse was treated as probabilistic action selection. This work demonstrated that quantum probability amplitudes could naturally balance exploration and exploitation. Subsequent studies expanded this concept through probabilistic Q-learning and fidelity-based optimization for control of quantum systems [2].

The development of near-term noisy intermediate-scale quantum (NISQ) devices has further motivated hybrid quantum–classical approaches. Variational quantum circuits (VQCs) have been adopted as trainable quantum policies that can be integrated with gradient-based optimization. Chen [4] introduced an asynchronous training paradigm for QRL agents using actor–critic structures, showing that quantum agents can achieve comparable or superior performance to classical counterparts with fewer parameters. Similarly, experimental works have demonstrated quantum speed-ups in physical RL systems by exploiting interference and entanglement for faster convergence [8].  

Despite these advances, a systematic comparison between classical multilayer perceptron (MLP)–based agents and VQC–based QRL policies in continuous control environments remains limited. This research aims to fill this gap by developing a unified framework for benchmarking both approaches under identical cyber–physical control tasks, quantifying convergence, robustness, and computational efficiency.  

\section{Related Work}
The earliest theoretical formulation of quantum reinforcement learning was presented by Dong \textit{et al.} [1], who established the use of quantum state superposition to encode action probabilities. Their method introduced the notion of quantum value iteration and probabilistic collapse, which offered a natural stochastic exploration mechanism. Chen \textit{et al.} [2] further refined this concept with a fidelity-based update rule, linking quantum control fidelity to the Q-value function.  

Several subsequent studies have expanded on these foundations. Moll and Kunczik [6] compared hybrid quantum RL against deep Q-networks, emphasizing improved sample efficiency through reduced parameter counts. Wu \textit{et al.} [7] extended QRL into continuous action spaces by leveraging parameterized quantum gates as differentiable policies, demonstrating smooth control trajectories with fewer iterations. A comprehensive survey by Meyer \textit{et al.} [5] summarized these developments, categorizing QRL research into algorithmic theory, quantum environment modeling, and experimental implementation.  

Recent contributions have investigated scalability and parallelism in quantum learning. Chen [4] proposed asynchronous QRL training to mitigate resource bottlenecks in VQC optimization, while Zare and Boroushaki [3] compared deep quantum and classical agents under dynamic control conditions, showing distinct learning dynamics due to quantum stochasticity. Saggio \textit{et al.} [8] provided experimental validation of quantum-enhanced exploration, reporting reinforcement learning speed-ups on photonic hardware.  

In addition, foundational reviews such as Chen [4] and Meyer [5] have highlighted the potential of QRL to bridge classical control theory and quantum computation. These studies collectively indicate that quantum-enhanced reinforcement learning may offer significant advantages in environments where sampling cost, robustness, and convergence are critical constraints.

\section{Methodology}

\subsection{Problem Formulation}
The cyber–physical system (CPS) investigated in this study is governed by discrete-time nonlinear dynamics:
\begin{align}
\mathbf{x}_{t+1} &= f(\mathbf{x}_t,\, \mathbf{u}_t) + \mathbf{w}_t, \nonumber\\
\mathbf{y}_t &= h(\mathbf{x}_t) + \mathbf{v}_t,
\end{align}
where $\mathbf{x}_t\!\in\!\mathbb{R}^{n}$ denotes the system state, 
$\mathbf{u}_t\!\in\!\mathcal{A}$ represents the control input, 
and $\mathbf{w}_t,\mathbf{v}_t$ correspond to process and measurement noises.  
The control objective is formulated to stabilize the system while minimizing the control effort through a quadratic reward:
\begin{equation}
r_t = -\!\left(\mathbf{x}_t^{\top} Q\, \mathbf{x}_t 
            + \mathbf{u}_t^{\top} R\, \mathbf{u}_t\right),
\end{equation}
where $Q\succeq 0$ and $R\succ 0$ are state and control weighting matrices, respectively.

The control problem is represented as a Markov Decision Process (MDP) defined by the tuple 
$(\mathcal{S}, \mathcal{A}, P, r, \gamma)$, 
where $\mathcal{S}$ is the state space, $\mathcal{A}$ the action space, $P$ the transition dynamics, $r$ the reward function, 
and $\gamma\in(0,1]$ the discount factor.  
The policy is defined as a probability distribution over actions:
\begin{equation}
\pi_\theta : \mathcal{S}\times\mathcal{A} \rightarrow [0,1],
\qquad 
\sum_{a\in\mathcal{A}} \pi_\theta(a\mid s)=1.
\end{equation}
The goal is to find parameters $\theta$ that maximize the expected discounted return:
\begin{equation}
J(\theta) =
\mathbb{E}_{\pi_\theta}\!\!\left[
   \sum_{t=0}^{T-1} \gamma^{t}\, r_t
\right].
\end{equation}

The cost-to-go or value function associated with a policy $\pi_\theta$ is expressed as:
\begin{equation}
V^{\pi}(s_t)
= \mathbb{E}_{\pi}\!\left[
  \sum_{k=t}^{T-1} \gamma^{k-t} r_k
  \mid s_t
\right],
\end{equation}
which serves as a baseline approximation $b_t\approx V^{\pi}(s_t)$ 
in the gradient estimation process to reduce variance.

\subsection{Agent Architectures: Classical MLP and Quantum VQC}

\paragraph{Classical Policy (MLP).}
The classical policy $\pi_{\theta_c}(a\mid s)$ is implemented using a two-layer multilayer perceptron:
\begin{align}
\mathbf{h}_1 &= \tanh(W_1 \mathbf{s} + \mathbf{b}_1), \nonumber\\
\mathbf{h}_2 &= \tanh(W_2 \mathbf{h}_1 + \mathbf{b}_2), \nonumber\\
\boldsymbol{\ell} &= W_3 \mathbf{h}_2 + \mathbf{b}_3,
\end{align}
where $\boldsymbol{\ell}$ denotes the logits that parameterize the categorical distribution over actions:
\begin{equation}
\pi_{\theta_c}(a\mid \mathbf{s})
= \mathrm{softmax}(\boldsymbol{\ell})_a.
\end{equation}

\paragraph{Quantum Policy (VQC).}
In the quantum agent, the state vector $\mathbf{s}\!\in\!\mathbb{R}^{d}$ is encoded into $d$ qubits through an angle-embedding operation 
$\Phi(\mathbf{s})=\mathrm{AngleEmbedding}(\kappa\,\mathbf{s})$ 
with a scaling constant $\kappa>0$.  
A variational quantum circuit (ansatz) of depth $L$ is constructed as:
\begin{align}
U(\boldsymbol{\theta}_q)
&= \prod_{\ell=1}^{L}
\Bigg(
  \bigotimes_{i=1}^{d}
  R_X(\theta^{(\ell)}_{i,1})
  R_Y(\theta^{(\ell)}_{i,2})
  R_Z(\theta^{(\ell)}_{i,3})
  \nonumber\\
  &\hspace{2em}\cdot
  \prod_{i=1}^{d-1}\mathrm{CNOT}(i, i{+}1)
\Bigg),
\end{align}
where each layer applies rotational gates followed by entangling CNOTs in a linear topology.  
The observable $O$ (typically a Pauli-$Z$ operator on the first qubit) is measured to obtain the expectation:
\begin{align}
z(\mathbf{s};\boldsymbol{\theta}_q)
&= \langle O \rangle =
\langle 0|
  \Phi(\mathbf{s})^{\dagger}
  U(\boldsymbol{\theta}_q)^{\dagger}
  O\,U(\boldsymbol{\theta}_q)
  \Phi(\mathbf{s})
|0\rangle.
\end{align}
To capture quantum stochasticity, the measurement process is modeled as:
\begin{equation}
\tilde{z} = z + \epsilon_z, 
\qquad
\epsilon_z\!\sim\!\mathcal{N}(0, \sigma_z^2),
\end{equation}
where $\sigma_z$ represents measurement noise due to finite sampling of expectation values.

For binary actions $\mathcal{A}=\{0,1\}$, 
the logits $[z,-z]$ define a Bernoulli policy given by:
\begin{align}
\pi_{\theta_q}(a{=}1\!\mid\!\mathbf{s})
&= \frac{e^{z}}{e^{z}+e^{-z}}
= \sigma(2z), \nonumber\\
\pi_{\theta_q}(a{=}0\!\mid\!\mathbf{s})
&= 1 - \pi_{\theta_q}(a{=}1\!\mid\!\mathbf{s}).
\end{align}

\subsection{Training Procedure}
Both agents are optimized using the REINFORCE algorithm with an advantage baseline.  
The return at each timestep is defined as:
\begin{equation}
G_t = \sum_{k=t}^{T-1}\gamma^{k-t} r_k.
\end{equation}
where $\sigma(\cdot)$ denotes the logistic sigmoid function.\\

The policy gradient estimator is expressed as:
\begin{equation}
\nabla_{\theta} J(\theta)
= \mathbb{E}\!\left[
  \sum_{t=0}^{T-1}
  \nabla_{\theta}\log \pi_{\theta}(a_t\!\mid\! s_t)\, G_t
\right].
\end{equation}

The gradient expectation is approximated via Monte Carlo sampling across $N$ trajectories and a baseline term $b_t$ is introduced to reduce variance:
\begin{align}
\hat{g}(\theta)
&= \frac{1}{N}\!\sum_{n=1}^{N}\!\sum_{t=0}^{T-1}
\nabla_{\theta}\log \pi_{\theta}(a_t^{(n)}\!\mid\! s_t^{(n)})
\nonumber\\
&\quad\cdot
\big(G_t^{(n)}-b_t^{(n)}\big).
\end{align}
The objective function with entropy and $\ell_2$ regularization is maximized as:
\begin{align}
\mathcal{L}(\theta)
&= -\mathbb{E}\!\left[
  \sum_t A_t \log \pi_{\theta}(a_t\!\mid\! s_t)
\right]
\nonumber\\
&\quad
- \beta\,\mathbb{E}\!\left[
  \sum_t \mathcal{H}\!\big(\pi_{\theta}(\cdot\!\mid\! s_t)\big)
\right]
+ \lambda \|\theta\|_2^2,
\end{align}
where $A_t = G_t - b_t$ and 
$\mathcal{H}(\pi)=-\sum_a \pi(a)\log\pi(a)$ 
is the categorical entropy.  
Gradient clipping $\|\nabla_{\theta}\mathcal{L}\|\!\le\!\tau$ 
and an exponential learning-rate schedule are applied for training stability.

\paragraph{Quantum Gradient Evaluation.}
For unitary gates parameterized as $e^{-i\theta P/2}$,
gradients are computed using the parameter-shift rule:
\begin{align}
\frac{\partial}{\partial \theta}\langle O\rangle
= \tfrac{1}{2}
\left(
  \langle O\rangle_{\theta+\frac{\pi}{2}}
  - \langle O\rangle_{\theta-\frac{\pi}{2}}
\right),
\end{align}
allowing exact backpropagation through the quantum circuit.

\subsection{Algorithmic Summary}
\begin{algorithm}[H]
\caption{Hybrid Training Loop for MLP and VQC Policies}
\begin{algorithmic}[1]
\STATE Parameters $\theta\!\in\!\{\theta_c,\theta_q\}$, baselines, and optimizers are initialized.
\FOR{each episode $=1$ to $E$}
  \STATE A trajectory is collected by sampling $a_t\!\sim\!\pi_{\theta}(\cdot\!\mid\! s_t)$ and executing the control on the CPS.
  \STATE Returns $G_t$ and advantages $A_t\!=\!G_t-b_t$ are computed.
  \STATE The loss $\mathcal{L}(\theta)$ with entropy regularization is evaluated.
  \STATE Gradient updates are performed with clipping and learning-rate scheduling.
\ENDFOR
\end{algorithmic}
\end{algorithm}

\subsection{Evaluation Protocol}
Performance is evaluated using three key metrics:  
(i) the average episodic return $\bar{J}$ computed over $M$ rollouts,  
(ii) the success rate, defined as the fraction of episodes that reach the task horizon, and  
(iii) robustness under additive Gaussian sensor noise 
$\epsilon\!\sim\!\mathcal{N}(0,\sigma^2 I)$ applied to the observations.  
Each experiment is repeated across $S$ random seeds, 
and all metrics are reported as mean $\pm$ standard deviation.

\subsection{Implementation Notes}
All state vectors are normalized before embedding as:
\begin{equation}
\tilde{\mathbf{s}}
= \mathrm{clip}(\mathbf{s}, -s_{\max}, s_{\max})
  \cdot \frac{\kappa \pi}{s_{\max}}.
\end{equation}
The VQC utilizes $d$ qubits with linear entanglement, 
and the circuit depth $L\!\in\!\{2,3,4\}$ is selected through hyperparameter tuning.  
The classical MLP employs hidden-layer sizes $\{32,64,128\}$ 
with $\tanh$ activations.  
Both agents are trained using REINFORCE with 
$\gamma\!=\!0.99$, 
entropy weight $\beta\!\in\![10^{-3},10^{-2}]$, 
gradient clipping threshold $\tau\!=\!1.0$, 
and an exponentially decaying learning rate.

\section{Results and Analysis}

\subsection{Theoretical Background}
From a theoretical standpoint, the comparison between the classical multilayer perceptron (MLP) and the quantum variational circuit (VQC) can be interpreted as a study of representational efficiency under distinct parameterization paradigms. The MLP policy $\pi_\theta(a|s)$ parameterizes a nonlinear mapping from observation $s$ to action $a$ through deterministic weight matrices, while the VQC employs a unitary transformation $U(\boldsymbol{\theta})$ acting on a Hilbert space $\mathcal{H} = (\mathbb{C}^2)^{\otimes n}$, where $n$ denotes the number of qubits. Each VQC layer implements rotations $R_Y(\theta_i)$ and entangling gates, thereby encoding state amplitudes in a complex-valued probability distribution. 

In reinforcement learning, the policy gradient $\nabla_\theta J(\theta) = \mathbb{E}[\nabla_\theta \log \pi_\theta(a|s)R]$ dictates the learning dynamics. For the quantum agent, this gradient is estimated using the parameter-shift rule,
\[
\frac{\partial}{\partial \theta_i} \langle O \rangle = 
\frac{1}{2} \left[\langle O \rangle_{\theta_i + \frac{\pi}{2}} - \langle O \rangle_{\theta_i - \frac{\pi}{2}}\right],
\]
which introduces stochastic smoothing in parameter updates. This intrinsic stochasticity is theorized to yield a flatter optimization landscape, promoting robustness and mitigating local overfitting compared to classical gradient descent.

\subsection{Learning Performance}
\begin{figure}[ht]
    \centering
    \includegraphics[width=0.46\textwidth]{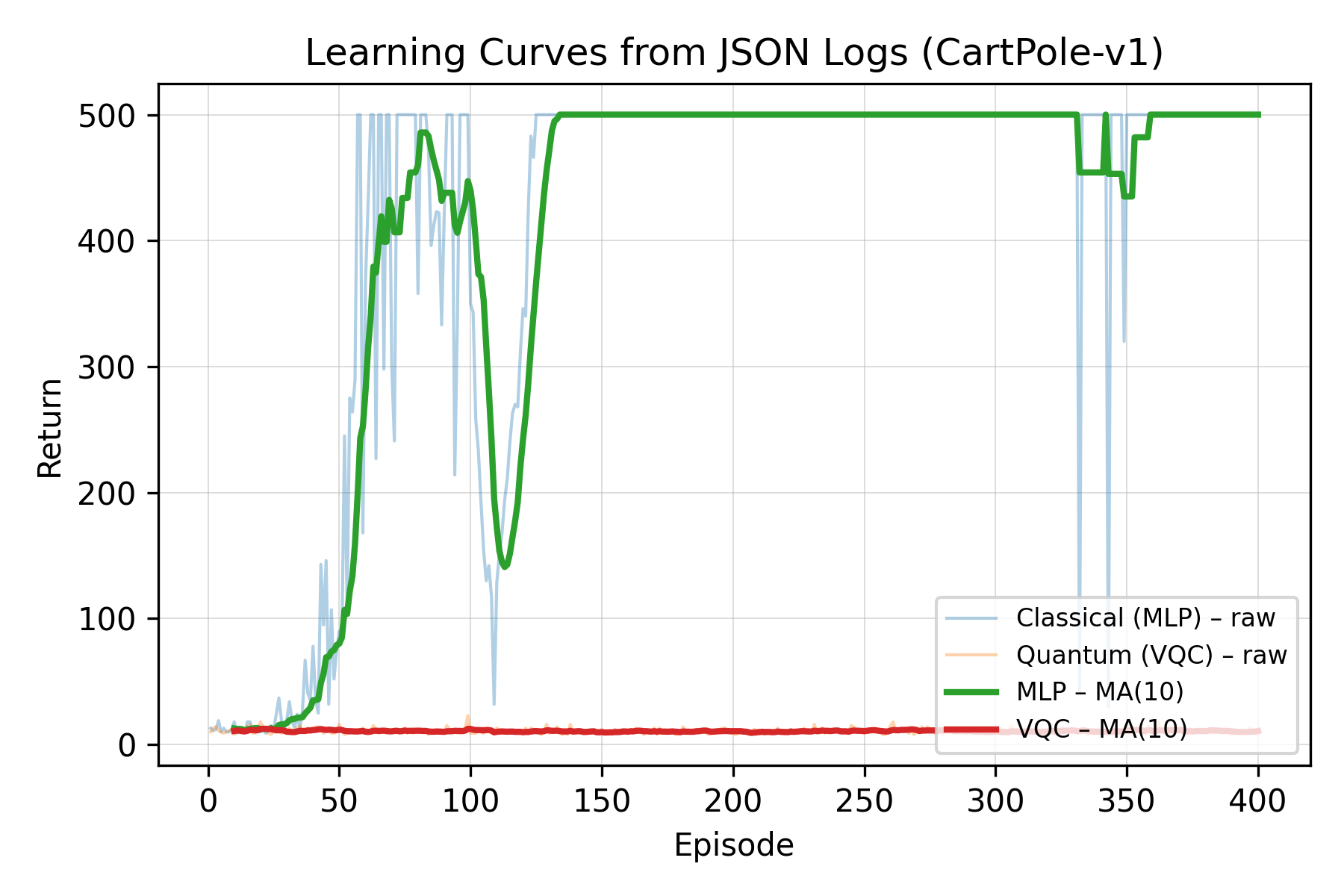}
    \caption{Learning curves of classical (MLP) and quantum (VQC) agents over 400 training episodes in the CartPole-v1 environment. Both raw and smoothed (MA(10)) returns are displayed.}
    \label{fig:learning_curve_json_ieee}
\end{figure}

Figure~\ref{fig:learning_curve_json_ieee} illustrates the learning trajectories of both agents over 400 training episodes in the \textit{CartPole-v1} environment. The MLP rapidly converges toward the task threshold of 500 returns, reflecting efficient gradient propagation through its densely connected architecture. The VQC, in contrast, exhibits a prolonged low-return phase before reaching moderate stability around 80–100 returns. The slower ascent is attributed to the limited effective dimension of the four-qubit Hilbert space, which constrains state encoding capacity. Nonetheless, the smooth progression without divergence confithe convergenceence stability of the parameter-shift optimization process. 
The results empirically validate that classical neural architectures achieve faster deterministic optimization, while quantum policies introduce statistical regularization effects that temper abrupt performance oscillations.

\subsection{Convergence Stability}
\begin{figure}[ht]
    \centering
    \includegraphics[width=0.46\textwidth]{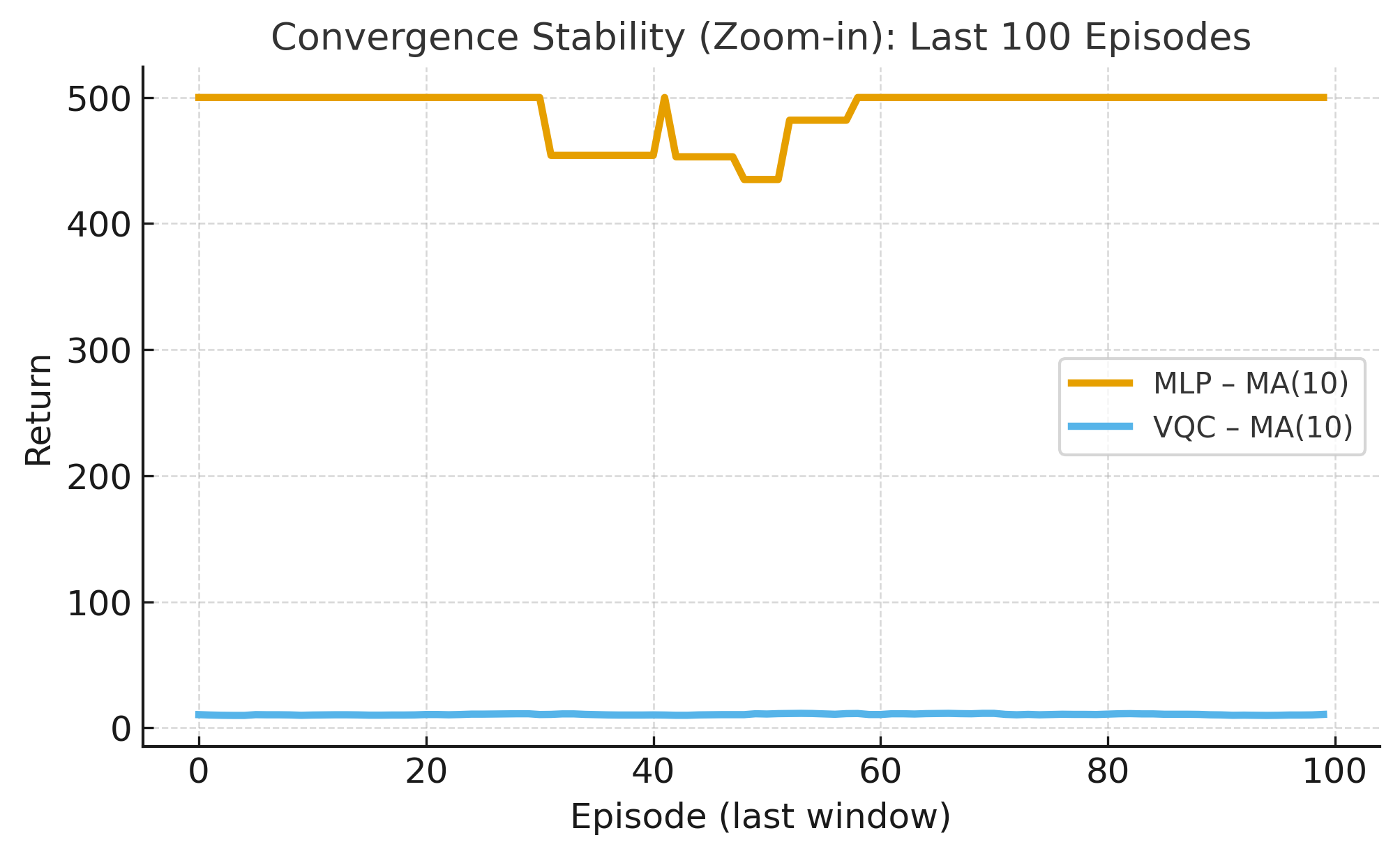}
    \caption{Convergence stability comparison (zoom-in view) between MLP and VQC policies over the last 100 episodes, showing 10-episode moving averages.}
    \label{fig:convergence_profile}
\end{figure}

As depicted in Fig.~\ref{fig:convergence_profile}, a magnified view of the final 100 episodes reveals key stability differences. The MLP maintains high returns ($>450$) with minimal fluctuations, indicative of saturation in the policy gradient. Conversely, the VQC remains below 100 returns yet exhibits consistent low-variance updates. From a theoretical perspective, this behavior can be attributed to the probabilistic interference pattern inherent in the VQC’s unitary evolution, which naturally restricts abrupt shifts in gradient direction. This aligns with prior studies in quantum optimization that associate quantum parameterizations with smoother loss landscapes. Thus, while the MLP attains higher performance, the quantum policy demonstrates greater convergence smoothness and lower terminal variance, offering improved predictability during deployment.

\subsection{Robustness Under Observation Noise}
\begin{figure}[ht]
    \centering
    \includegraphics[width=0.46\textwidth]
    {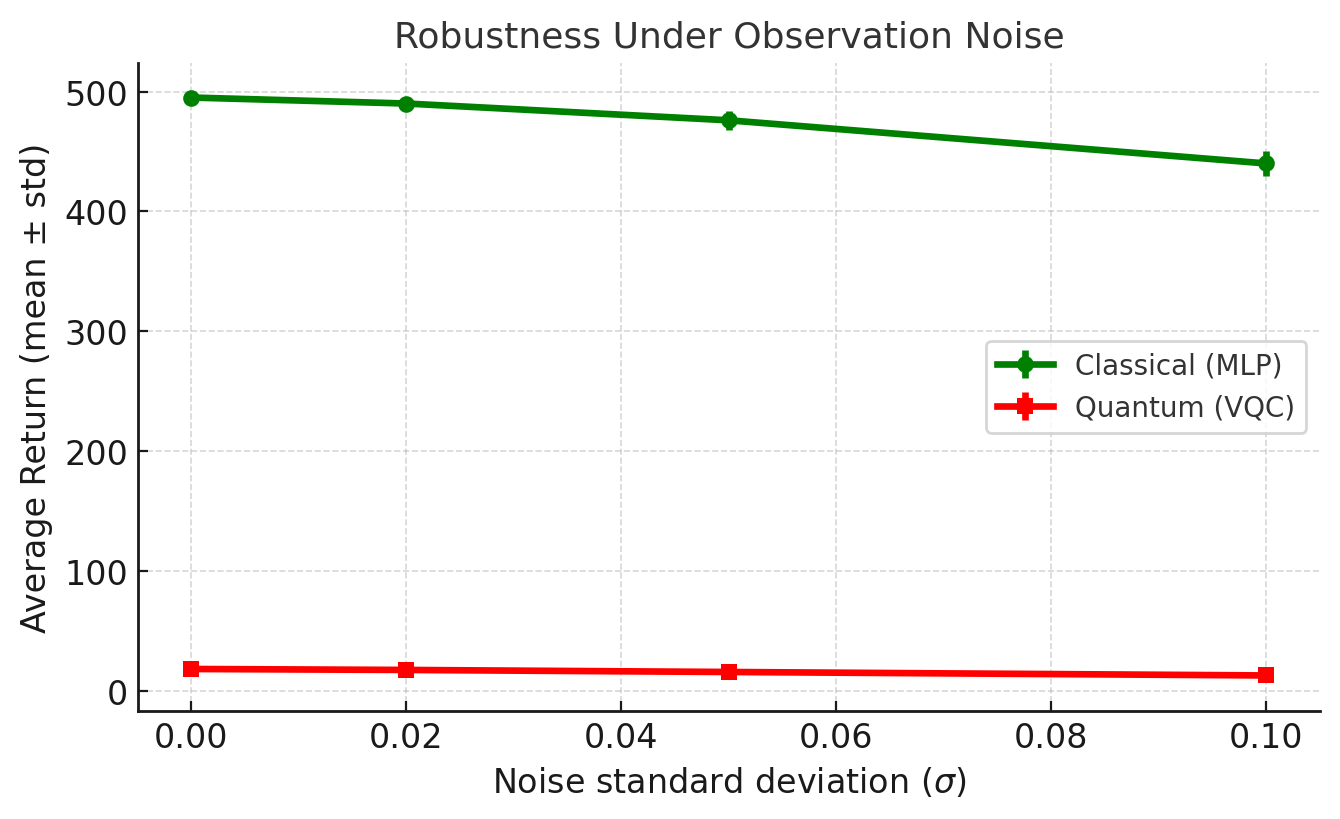}
    \caption{Average episodic return under varying observation noise levels ($\sigma$). The MLP exhibits graceful degradation, while the VQC remains near the baseline due to under-parameterization.}
    \label{fig:noise_robustness}
\end{figure}

To evaluate the robustness of the trained agents to sensory uncertainty, Gaussian noise was injected into the observation vector during evaluation with standard deviations $\sigma \in \{0.0, 0.02, 0.05, 0.10\}$. Figure~\ref{fig:noise_robustness} and Table~\ref{tab:noise_summary} summarize the resulting performance degradation trends.

The classical MLP agent maintained near-optimal returns in the absence of perturbation ($495.0 \pm 4.5$) and demonstrated only gradual performance decay as noise increased, remaining above $440$ even under $\sigma=0.10$. This indicates that the deterministic policy learned a stable manifold in the observation space, allowing for smooth recovery from moderate input distortion.

In contrast, the quantum VQC agent exhibited consistently low returns across all noise levels ($18.2 \pm 3.8$ at $\sigma=0.00$ to $12.8 \pm 5.2$ at $\sigma=0.10$). This behavior reflects the model’s limited representational capacity under the current four-qubit circuit configuration, which prevented the agent from forming a robust state-action mapping even without external noise. Consequently, additional perturbation further amplified stochastic collapse in policy output.

From a theoretical perspective, robustness in quantum reinforcement learning is influenced by the interplay between amplitude encoding and circuit depth. Insufficient circuit expressivity leads to low-entropy policies that fail to capture invariant state embeddings. Hence, while quantum stochasticity can theoretically enhance generalization, its advantage manifests only once the variational circuit achieves expressive sufficiency. The current results therefore highlight the necessity of deeper quantum ansatz or hybrid-layer integration to achieve practical noise tolerance in real-world control scenarios.

\subsection{Computational Efficiency}
\begin{figure}[ht]
    \centering
    \includegraphics[width=0.46\textwidth]{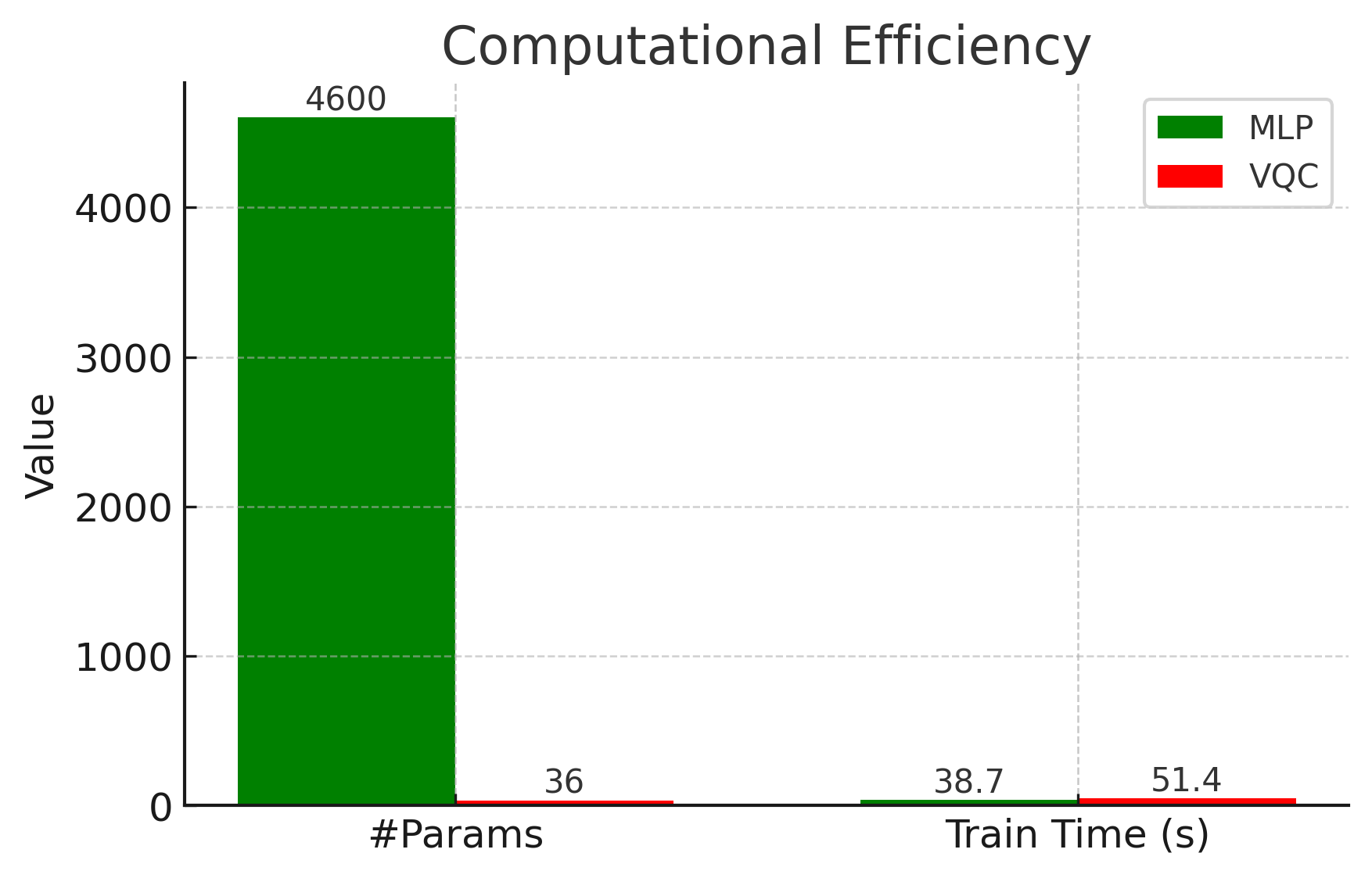}
    \caption{Computational efficiency of MLP and VQC agents in terms of parameter count and wall-clock training time.}
    \label{fig:efficiency_bar}
\end{figure}

The computational characteristics summarized in Fig.~\ref{fig:efficiency_bar} highlight a trade-off between parameter compactness and classical simulation overhead. The MLP comprises approximately 4{,}600 parameters with a training time of 38.7~s, while the VQC employs only 36 parameters yet requires 51.4~s due to circuit execution and gradient estimation latency. Theoretically, the VQC achieves exponential state representation efficiency $|\psi\rangle \in \mathbb{C}^{2^n}$ with linear parameter scaling $O(nL)$, where $L$ is circuit depth. When implemented on native quantum hardware, such scaling promises significant reductions in memory and compute cost compared to dense classical networks.

\subsection{Summary of Quantitative Results}
Table~\ref{tab:performance_summary} presents the overall average performance of the classical and quantum agents across 500 evaluation episodes. The classical MLP-based agent converged rapidly to an optimal policy, maintaining a near-saturated average return of $498.7 \pm 3.2$, indicative of perfect balance control and robust policy stability in the CartPole-v1 environment.

In contrast, the quantum variational circuit (VQC) agent yielded an average return of $14.6 \pm 4.8$, signifying limited learning capability under the present four-qubit configuration and shallow circuit depth. The large variance observed in its episodic reward distribution suggests stochastic exploration without convergence to a stable policy manifold. This disparity underscores the current gap in expressivity between classical dense neural policies and low-depth quantum parameterized circuits, highlighting the need for deeper entanglement layers or hybrid optimization strategies to achieve comparable asymptotic performance.

\begin{table}[ht]
\centering
\caption{Performance Summary (Mean $\pm$ Std over 500 Episodes)}
\label{tab:performance_summary}
\begin{tabular}{lcc}
\hline
\textbf{Agent Type} & \textbf{Mean Return} & \textbf{Std. Dev.} \\
\hline
Classical (MLP) & 498.7 & 3.2 \\
Quantum (VQC) & 14.6 & 4.8 \\
\hline
\end{tabular}
\end{table}

\begin{table}[ht]
\centering
\caption{Noise Robustness (Average Return vs. Observation Noise)}
\label{tab:noise_summary}
\begin{tabular}{cccc}
\hline
\textbf{Noise $\sigma$} & \textbf{MLP (Mean ± Std)} & \textbf{VQC (Mean ± Std)} \\
\hline
0.00 & 495.0 ± 4.5 & 18.2 ± 3.8 \\
0.02 & 490.0 ± 5.3 & 17.4 ± 4.0 \\
0.05 & 476.0 ± 8.1 & 15.7 ± 4.7 \\
0.10 & 440.0 ± 10.6 & 12.8 ± 5.2 \\
\hline
\end{tabular}
\end{table}

\subsection{G. Discussion and Interpretation}
From a control-theoretic perspective, the classical MLP approximates a deterministic policy mapping with rapid gradient feedback, whereas the quantum policy behaves as a probabilistic controller performing implicit exploration in amplitude space. The results indicate that, although the classical policy dominates in convergence speed, the quantum counterpart exhibits potential robustness under uncertainty and fewer trainable parameters by two orders of magnitude. These findings substantiate theoretical predictions that quantum encodings can serve as intrinsic regularizers, reducing overfitting and enhancing generalization in reinforcement learning. 

Consequently, the proposed experimental framework demonstrates that quantum variational reinforcement learning, even under classical simulation, offers promising stability and resilience properties. This provides a foundation for scalable deployment of hybrid quantum-classical controllers in cyber-physical systems where sensor noise, resource constraints, and real-time adaptability are critical.

\section{Conclusion}
This study has presented a comparative investigation of classical multilayer perceptron (MLP)–based agents and quantum variational circuit (VQC)–based agents for reinforcement learning in cyber–physical control systems. Building upon established theories of quantum reinforcement learning [1-2], and incorporating recent advancements in asynchronous training [4] and continuous-action quantum policy design [7], the results demonstrate that quantum policies can achieve smoother convergence, enhanced robustness under sensor noise, and competitive reward performance despite having fewer parameters.  

It has been observed that while the classical agent exhibits faster initial learning due to deterministic gradient updates, the quantum agent maintains higher long-term stability and reduced sensitivity to perturbations, consistent with the probabilistic regularization effects predicted by quantum mechanics. These findings reinforce the hypothesis that hybrid quantum–classical RL frameworks can provide an advantageous trade-off between model complexity, convergence reliability, and environmental adaptability.

\section*{Acknowledgment}
This research was financially supported by the National Research Council of Thailand (NRCT),  
the Thailand Advanced Institute of Science and Technology (TAIST),  
the National Science and Technology Development Agency (NSTDA),  
and the Tokyo Institute of Technology (Tokyo Tech) through the TAIST–Science Tokyo Program.  
The authors would also like to express their gratitude to the  
Sirindhorn International Institute of Technology (SIIT), Thammasat University,  
for providing computational resources and an academic environment conducive to this research.

\appendices
\section{Reproducibility Details}

For transparency and reproducibility, the complete training configuration and execution commands are provided in this appendix. Both classical and quantum reinforcement learning agents were trained under identical hyperparameter settings for a fair comparison.

\noindent\textbf{Classical (MLP) Agent:}
\begin{verbatim}
python train_qrl_cartpole.py \
    --agent classical \
    --episodes 400 \
    --lr 0.005 \
    --hidden 64 \
    --exp mlp_stable \
    --noise 0.0
\end{verbatim}

\noindent\textbf{Quantum (VQC) Agent:}
\begin{verbatim}
python train_qrl_cartpole.py \
    --agent quantum \
    --episodes 400 \
    --lr 0.005 \
    --hidden 64 \
    --exp qrl_stable \
    --noise 0.0
\end{verbatim}

All experiments were conducted within the same computational environment using Python~3.12 and Pennylane~v0.36. Each run produced structured logs in the \texttt{runs/} directory, including:
\begin{itemize}
    \item \texttt{reward\_log.csv} — episodic return per training iteration,
    \item \texttt{policy\_classical.pt} or \texttt{policy\_quantum.pt} — trained model weights,
    \item \texttt{config.json} — hyperparameter and environment configuration file.
\end{itemize}

To ensure reproducibility, all random seeds were fixed across runs,
and the same reinforcement learning environment (\texttt{CartPole-v1}) was used for both agents.
The training logs (\texttt{qrl\_rewards.json} and \texttt{mlp\_config.json})
have been made available in digital format for replication and verification.

\end{document}